\documentclass[twoside]{article}
\usepackage{graphicx}
\usepackage{fancyhdr}
\usepackage{multicol}
\usepackage{styl-ap}

\begin{document}
\title{Recurrent Novae -- A Review}
\author{Koji Mukai\work{1,2}}
\workplace{CRESST and X-ray Astrophysics Laboratory, NASA/Goddard Space Flight Center, Greenbelt, MD 20771, USA
\next
Department of Physics, University of Maryland, Baltimore County, 1000 Hilltop Circle, Baltimore, MD 21250, USA}
\mainauthor{Koji.Mukai@nasa.gov}
\maketitle

\begin{abstract}%
In recent years, recurrent nova eruptions are often observed very
intensely in wide range of wavelengths from radio to optical to X-rays.
Here I present selected highlights from recent multi-wavelength
observations. The enigma of T Pyx is at the heart of this paper.
While our current understanding of CV and symbiotic star evolution
can explain why certain subset of recurrent novae have high accretion
rate, that of T~Pyx must be greatly elevated compared to the evolutionary
mean.  At the same time, we have extensive data to be able to estimate
how the nova envelope was ejected in T~Pyx, and it turns to be a
rather complex tale. One suspects that envelope ejection in recurrent
and classical novae in general is more complicated than the textbook
descriptions. At the end of the review, I will speculate that these
two may be connected.
\end{abstract}


\begin{multicols}{2}

\section{Introduction}

Nova eruptions are understood to be powered by thermonuclear runaway
(TNR) on the surface of accreting white dwarfs.  Hundreds of objects
in the Galaxy have been seen to experience one nova eruption: these
are called classical novae (CNe).  Recurrent novae (RNe) are objects
that have been seen to experience multiple nova eruptions.  There are
currently 10 confirmed RNe in the Galaxy.  Between 10$^{-6}$ and
10$^{-4}$ M$_\odot$ of hydrogen rich material needs to be accreted
to reach the critical temperature and density required for TNR.
The critical mass is lower for more massive white dwarfs with higher
gravity.  Therefore, we expect RNe to contain near Chandrasekhar mass
white dwarf accreting at a high rate.  This makes RNe candidate
progenitors of Type Ia supernova.  For this reason, and because
the recurrent nature of these objects allows studies that one cannot
undertake for CNe, RNe have become the subject of intensive study.

It is impossible to present a comprehensive review of RNe in the
space allotted; for that, the readers are referred to Schaefer
(2010) and Anupama (2013).  In this review, I will present selected
highlights from multi-wavelength campaigns on recent RN outbursts,
highlighting the work of the E-Nova
collaboration\footnote{https://sites.google.com/site/enovacollab/}.
I will also include results on several CNe: some of these system may
be unrecognized or unconfirmed RNe, and others provide a useful
comparison.  I will also present some quiescent observations.
I will discuss implications on the white dwarf mass, the ejecta mass,
the quiescent accretion rate,and the evolutionary scenarios for RNe and CNe.

\subsection{X-ray Bursts: a cautionary tale}

Although RNe provide a unique opportunity to compare multiple
nova episodes and possibly to compare accreted vs. ejected mass,
only a handful of eruptions are observed for each system.  This
is in stark contrast to the studies of X-ray bursts, which are
TNRs on accreting neutron stars.  For example, Linares et al.
(2012) studied 398 X-ray bursts detected from the transient
X-ray binary in the globular cluster, Terzan 5, as the accretion
changed by a factor of $\sim$5.  This allowed these authors to
study the relationship between the persistent luminosity,
the burst recurrence time and the burst fluence, and thereby
test the theory of TNR on neutron stars. Unfortunately, analogous
tests have not been possible yet in the case of RNe.

Yet, even in the case of X-ray bursts, puzzles remain
(Galloway et al. 2008).  One is the burst oscillations observed
during the decay.  The drifting period of burst oscillations
reflect the spin period of the neutron star atmosphere,
which changes as the atmosphere expands and then contracts
during the course of a burst.  The presence of the oscillations
during the decay, however, requires inhomogeneous burning
over the neutron star surface, even though one might expect
uniform burning at this stage.  The other is that pairs of
bursts can occur with very short ($<$10 min) recurrence times,
much too short to have accreted sufficient fuel for a new
burst, judging by the persistent X-ray luminosity.  This
requires a reservoir of unburnt fuel on or very near the
neutron star surface.

Thus, our theoretical understanding of X-ray bursts appears
incomplete.  It may well be that the current theories of
nova outbursts are equally incomplete regarding, e.g., the
recurrence times of RNe.

\section{Selected Recent Results}

\subsection{Ejecta Geometry}

Montez et al. (in preparation) have detected extended X-ray
emission in the {\sl Chandra\/} observations of RS Oph obtained
in 2009 and 2011.  These structures are well-separated from
the central X-ray source in the E-W direction, and were seen
to expand from 2009 to 2011. This X-ray emitting bipolar
outflow appears to follow the same angular expansion curve
inferred for radio and {\sl Hubble Space Telescope (HST)\/}
bipolar structures observed earlier.  The implied current expansion
velocity is very high (of order 4,000 km\,s$^{-1}$. One possible
origin of the bipolar flow is that RS Oph produced a true,
well-collimated, jet near the time of nova eruption.  Another is
that an initially spherical ejecta encountered an equatorial
torus and slowed down except in the polar directions.  Since
RS Oph is an RN in a symbiotic binary, the wind of the giant
mass donor is a potential source of such a torus (Mohamed
et al. 2013).

However, similar shaping of the ejecta might also occur in
cataclysmic variables (CVs), with a Roche-lobe filling mass
donor on or near the main sequence.  In a series of simulations
of the 2010 eruption of U Sco by Drake \& Orlando (2010), the
accretion disk is destroyed by the blast wave.  This interaction
causes the ejecta to expand away from the orbital plane.  One
particular simplifying assumption used by these authors, that
of a uniform density disk, is a cause for concern, and independent
simulations are needed to confirm their results in general.
Nevertheless, the possibility that disk-blast wave interactions
create bipolar outflow should be kept in mind for all novae,
whether the underlying binary is a symbiotic system or a CV.

The above-mentioned results on RS Oph and U Sco are both about
the outflow during the most recent outbursts of RNe, and may
apply to CNe as well.  In contrast, one type of study
unique to RNe is the analysis of light echoes produced by
ejecta from previous outbursts, as Sokoloski et al. (2013)
did for T Pyx.  The arrangement of the echo location on the sky
and the progression of echos from east to west suggest a ring-like
structure from a previous outburst.  The delay times for echoes
along the north-south axis suggest a distance of 4.8$\pm$0.5 kpc
for T Pyx.  Moreover, the time lags between different echoes
suggest that the ring is inclined $\sim$30--40$^\circ$ relative
to the plane of the sky.  This is most likely to reflect the
binary inclination, somewhat higher than values previously
inferred for T Pyx.  Regardless of the precise inclination
angle, the very fact that an equatorial ring was formed by
the ejecta is worth noting.

\subsection{Novae in Symbiotic Systems}

Four of the known Galactic recurrent novae are in symbiotic
binaries: RS Oph, T CrB, V745 Sco, and V3890 Sgr.  They are
all S type systems: they have a normal red giant mass donor,
an orbital separation of order 1 AU, and an orbital period
of order 1--2 years.  In the other subtype, the D (dusty)
type, the mass donor is an AGB star; the D type systems
have a much wider orbit than the S type systems.  Before
2010, all known TNR events in symbiotic systems were either
a very slow ``symbiotic novae,'' or very fast RNe (Mikolajewska
2008).  It is important to note that TNR can lead to a quasi-static
configuration without explosive mass loss in symbiotic novae.
Also noteworthy is the fact that the accretion rate is high
enough in the 4 S type symbiotic systems to produce RNe.
If we take 10$^{-7}$ M$_\odot$\,yr$^{-1}$ as the typical wind
mass loss rate of a normal red giant, then this implies
either Roche-lobe overflow or a very efficient mechanism
to capture the wind, such as wind Roch-lobe overflow
(Mohamed \& Podiadlowski 2010), although M giants in
symbiotic binaries may have higher mass-loss rates
(Seaquist \& Taylor 1990).

In March 2010, a D-type symbiotic system, V407 Cyg, became a nova.
It was noteworthy for being the first nova to be detected
as GeV $\gamma$-ray source with {\sl Fermi\/} LAT (Abdo
et al. 2010); it was the subject of an intensive multiwavelength
from radio to X-rays (Nelson et al. 2012; Chomiuk et al. 2012).
The X-rays were predominantly from the shock between the
nova blast wave and the wind of the Mira type mass donor;
interestingly, V407 Cyg became X-ray bright after the
GeV signal faded.  The thermal emission from the flash-ionized
AGB wind was the dominant source of radio signal.  While
we learned a lot about the nova event, we are left with
one important question: how often does V407 Cyg experience
nova eruptions?  Is it an unrecognized RN, or are the
eruptions much less frequent?

\subsection{Long Period CVs}

Darnley et al. (2012) proposed to classify novae into
red giant, sub-giant, and main sequence systems.  The
orbital periods of the ``sub-giant'' systems are in
the range 10 hrs to 6 days.  According to the Ritter
\& Kolb catalog Version 7.20 (Ritter \& Kolb 2003),
there are 46 systems (excluding one uncertain entry
in the catalog) in this orbital period range.  Of the
confirmed RNe, V894~CrA, U~Sco and CI~Aql belong to
this group, and a fourth, V2487~Oph, probably has an
orbital period in this range.

The evolution of such long-period CVs has not been
studied extensively to date, compared to those with
periods under 10 hrs, for which the basic framework
and much more have been established (Knigge et al. 2011).
CVs with similarly long orbital periods include several
novae not known to be recurrent (GK~Per, V1017~Sgr),
supersoft sources (e.g., MR~Vel, CAL~83), and V~Sge and
other systems that may be related to supersoft sources.
One possibility is that these systems are currently undergoing
thermal timescale mass transfer (TTMT; Schenker et al. 2002)
or did so in the past.  The evolution of CVs with P$>$10 hrs
should be studied further, and a search for additional
nova outbursts of systems like GK~Per may be worthwhile.

\subsection{Quiescent X-ray Observations}

\begin{myfigure}
\centerline{\resizebox{70mm}{!}{\includegraphics{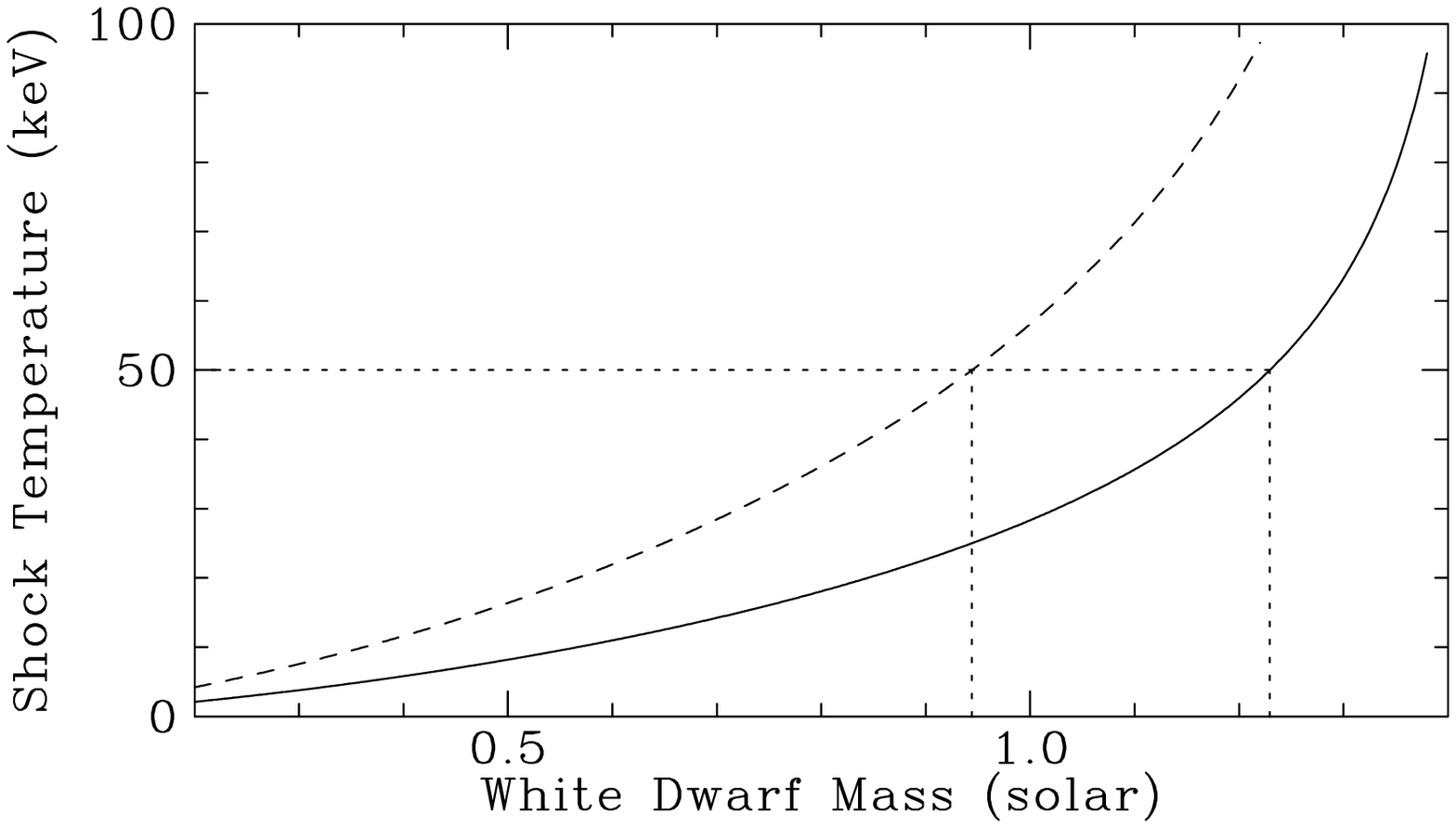}}}
\caption{Possible white dwarf mass of V2487~Oph for the magnetic
(dashed line) and the non-magnetic (solid line) cases}
\label{km-fig1}
\end{myfigure}

Of the known recurrent novae, T~CrB and V2487~Oph are
bright hard X-ray sources detected with {\sl INTEGRAL\/}
and with {\sl Swift\/} BAT (see, e.g., Baumgartner et al. 2013).
When such hard X-ray emission is detected, it can constrain
both the white dwarf mass and the accretion rate in quiescence.
In both magnetic and non-magnetic cases, hard X-rays are generated
when the supersonic accretion flow encounters the white dwarf surface
and is shock-heated.  The hot plasma must radiatively cool before
settling onto the white dwarf surface.  In the case of magnetic
CVs, the accretion flow is radial and its velocity is approximately
at the free-fall value (Aizu 1973).  The observed X-ray spectrum
is multi-temperature in nature, with shock temperatures typically
in the 10--50 keV range.  The spectral curvature in the hard
X-ray range is a reliable measure of the shock temperature,
and can be used to infer the white dwarf mass (see, e.g.,
Yuasa et al. 2010).  The situation may well be more complex
for the boundary layer of non-magnetic CVs, but the maximum
temperature is unlikely to exceed that expected for strong
shocks from Keplerian velocity, and quiescent dwarf novae
appear to follow the ``Keplerian strong shock'' relationship
(Byckling et al. 2010).

Some authors have speculated that V2487~Oph may contain a
magnetic white dwarf, based solely on its bright, hard X-ray
emission.  However, I measure a shock temperature of $\sim$50
keV using the BAT survey 8-channel spectrum, which corresponds
to either a $\sim$0.9 M$_\odot$ magnetic white dwarf or a
$>$1.24 M$_\odot$ non-magnetic white dwarf.  Given the RN nature
of this object, and given that repeated {\sl XMM-Newton\/}
observations have not revealed a spin modulation, the latter
interpretation seems more likely.

\subsection{High Mass Transfer Rate: Evolutionary or Temporary?}

Patterson et al. (2013) made the following simple prediction, based on
average secular mass accretion rate:  CVs above the period gap
should experience nova eruptions once every 10,000 years or so,
while those below the gap should recur once every 1 million year.
For these normal CVs to be a RN, the accretion rate needs to be
elevated by many orders of magnitudes above the secular mean.
This is in contrast to the long-period systems with a sub-giant
mass donor, where TTMT may drive a very high accretion rate.
Similarly, in symbiotic systems, the mass loss rate from the
donor is high enough, although the fraction that can be captured
by the white dwarf is highly uncertain. Let us now
consider the census of known classical and recurrent novae
of various types with the above expectations in mind.

Among symbiotic stars, most novae are extremely slow and often
referred to as symbiotic novae, the slowness suggestive of low-mass
white dwarfs.  The four
well-known symbiotic recurrent novae are all in S-type systems:
the short recurrence time and the high velocity of the ejecta both
suggest these to have massive white dwarfs. The bifurcation of novae
in symbiotic stars into two such extreme groups is very different from
the situation in CVs, and should be investigated further. Although
only a single outburst is known, the outburst properties of V407 Cyg
makes it a candidate RN in this context.  Since it is in a D type
symbiotic system, with a much greater binary separation, an estimate
of its quiescent accretion rate, when one is obtained, may tell us about
how white dwarfs accrete in symbiotic stars, not to mention providing
a clue as to its likely recurrence time.

U~Sco, CI~Aql, and V894~CrA are long-period systems; V2487~Oph
may belong in this group, although its orbital period has not
been determined yet.  The TTMT scenario points to the general
framework for why these systems can be RNe.  More research is
needed to understand why some long-period systems are SSS,
others RNe, and yet others CNe (as far as we know).

This leaves IM~Nor (in the period gap) and T~Pyx (below the gap)
in the orbital period range with very low secular accretion rate
and with only a small number of known classical novae (V1794~Cyg,
V~Per, ...).  It is interesting to note that, while many classical
novae are known above the gap (orbital period in the 3--10 hr range),
no RNe are known in this regime.  This could purely a matter of
small number statistics.  At the same time, it could be that the
mechanism elevating the accretion rate of T~Pyx and IM~Nor far
above the secular mean is much less effective for systems above
the gap.

\subsection{Multi-Wavelength Observations of T Pyx}

Multi-frequency radio monitoring of novae is a powerful technique
that allows us to estimate the total ejected mass in a relatively
simple manner, although complications often arise (see, e.g., Roy
et al. 2012 and references therein.)  The nova ejecta is initially
optically thick in the radio, thus the brightening traces the
angular expansion of the ejecta.  As the ejecta becomes optically
thin, first from the highest frequencies then progressively to
lower frequencies, this allows the amount of mass to be estimated,
as long as we have a handle on the temperature and clumping of
the ejecta.  At the same time, we know that early X-ray emission
from novae is likely due to shocks in the ejecta (e.g., O'Brien
et al. 1994).  With this mind, the E-Nova collaboration has begun
multi-wavelength observations of recent novae using the much
improved Karl G. Jansky VLA in the radio, and {\sl Swift\/} and
other observatories in the X-rays.

T~Pyx is one of the major targets of the E-nova collaboration.
The radio and X-ray results are presented by Nelson et al. (2014)
and Chomiuk et al. (2014), respectively.  T~Pyx was largely
undetected in the radio for the first $\sim$60 days since the
discovery of the 2011 outburst, then started to rise around
day $\sim$100.  It was also X-ray faint during the first
several months, and then started to rise slightly after the
onset of the radio rise.  The X-ray photons detected with
{\sl Swift\/} XRT are a mixture of optically thin emission from
the shocked shell and the supersoft emission from the still
nuclear burning surface.  It is difficult to disentangle the
two using short snapshot observations, hence it is difficult
to determine, e.g., the turn-on time of the supersoft emission.

In the optical, T Pyx remained near peak optical magnitude
for 2 or 3 months, depending on where one defines the peak
to have ended and decline to have begun. This implies a large
photospheric radius, perhaps of order 5$\times 10^7$ km
($\sim$ a third of an AU; assuming a blackbody with a temperature
of 10,000K, a distance of 4.8 kpc, and A$_V \sim$ 1.0, this radius
corresponds to a visual magnitude of V$\sim$7.9).  This is well
outside the central binary, yet it only takes of order 1 day for matter
traveling at 600 km\,s$^{-1}$ to reach this distance.  For a
shell at a distance of 1$\times 10^8$ km to have an optical
depth of 1, which implies a column density of order 3 g\,cm$^{-2}$
(assuming electron scattering dominates the opacity), the total mass
of the shell must be greater than $\sim 5.0 \times 10^{-7}$
M$_\odot$.  So the duration of optical peak implies either a continuous
ejection of $\sim 5 \times 10^{-7}$ M$_\odot$ per day for several months,
or that T~Pyx went into a quasi-static, red giant-like configuration
during the peak, and ejected the extended atmosphere with a significant
delay.  In the latter case, the total mass of the extended atmosphere
must be much greater than $5 \times 10^{-7}$ M$_\odot$, because it is
a filled sphere and not a thin shell.

Schaefer et al. (2013) presented the detailed visual light curve
of T~Pyx during the initial rise.  Until it reached V$\sim$7.7,
it can be modeled well assuming a uniform expansion of the ejecta,
then the observed brightness drops below this model.  If we equate
this instance with the time when the optical depth of the ejecta
dropped below 1.0, we infer that an initial shell of
$\sim 6 \times 10^{-7}$ M$_\odot$ was ejected.

Such a small ejection is easy to hide in the radio data, although
there is one detection on day 17 that could be interpreted as due
to this.  On the other hand, continuous mass ejection of
$\sim 5 \times 10^{-7}$ M$_\odot$ per day is difficult to reconcile
with the deep radio non-detections followed by rapid brightening
around day 60. Rather, the radio data are consistent with a prolonged
period of quasi-stationary configuration, and a delayed ejection of
a more massive (of order 10$^{-5}$ M$_\odot$) shell.  In addition,
if the latter system was ejected with a larger velocity, then we expect
shocked X-ray emission when it catches up with the initial ejecta.
The existing X-ray data are broadly consistent with such a picture.

\subsection{The Cause of Elevated Mass Transfer Rate}

For T~Pyx, we have a clear-cut case that the mass transfer rate is
elevated by several order of magnitudes above the evolutionary
mean (Gilmozzi \& Selvelli 2007). The same presumably applies to
IM~Nor as well.

Irradiation of the donor is often invoked as the explanation. However,
theorists have long concluded that this requires hard photons,
and therefore theoretical studies largely focus on X-ray binaries
(see, e.g., Hameury et al. 1986; King 1989).  To quote from Ritter (2000),
``Energy emitted in certain spectral ranges, as e.g., EUV radiation
and soft X-rays, is unlikely to reach the photosphere of the donor."
To reach the photosphere of K or M type dwarfs, irradiating flux
needs to be able to penetrate above 10$^{24}$ cm$^{-2}$ of column,
thus requiring strong flux above 10 keV.  Therefore, supersoft sources
or photospheric emission of otherwise very hot white dwarf only
irradiates the chromosphere and above, and not the photosphere.
The irradiation mechanism studied in above-mentioned papers cannot
work when the irradiating flux is in the form of soft X-ray and
EUV photons.

While V1500 Cyg is sometimes taken as an example of a system
that is experiencing enhanced accretion rate due to irradiation, this
is not necessarily the case. This is because one well-established
effect of irradiation is to increase the luminosity of the
existing structures, be it the secondary or the accretion disk.
In fact, according to Somers \& Naylor (1999), the elevated brightness
of V1500 Cyg, which is currently an asynchronous polar due to its
1975 Nova eruption, is due to an orbitally modulated component, not 
due to the spin modulated component. Since accretion luminosity
should be modulated on the spin period, we know that the extra light
is due to the irradiated face of the secondary. In this picture, the
reflection of the gradually decreasing post-nova white dwarf flux
explains the secular changes in the brightness of V1500 Cyg, without
invoking variable accretion rate.

Thomas et al. (2008) obtained phase-resolved K-band photometry
of old novae of various ages since outburst.  They were also
able to interpret their results without invoking changing
accretion rate. Rather, in their interpretation, variable irradiation,
and hence variable reflection, changes the brightness of the
existing structures, the accretion disk and the secondary.
At a minimum, this points out that an enhanced brightness is
insufficient to prove an enhanced accretion rate.  These studies
also suggest that irradiation by a post-nova does not lead to
enhanced mass transfer, although they do not yet constitute a
solid proof.

If not irradiation, what other mechanisms can potentially enhance
the mass transfer? Here I speculate that the post-nova common
envelope phase might be ultimately responsible, as follows.

The ring geometry of the ejecta of T~Pyx (\S 2.1) suggests that the
secondary plays a role in shaping the geometry of the ejecta. In
symbiotic systems, slow novae can stay in the ``plateau'' phase
for decades, whereas no such example is known among CVs, again
implying that the secondary plays a role in ejecting the nova
envelope. The potential role played by the common envelope system
was first pointed out by MacDonald (1980) and later studied
quantitatively by, e.g., Livio et al. (1990).  The general consensus
is that the common envelope phase can contribute to the ejection
of the nova envelope, but only if the ejecta is moving more slowly
than the orbital motion of the mass donor.  The multi-wavelength
data on T~Pyx indeed suggests that the envelope may have been
in a quasi-stationary configuration for 2--3 months, as it is
for decades in slow symbiotic nova. The possibility that many
novae in CVs may not be able to eject the bulk of the envelope,
if it were not for the common-envelope phase, should therefore
be investigated.

If common envelope phase is an important factor in the envelope
ejection process, then this implies that the binary must have
lost angular momentum.  While the orbital period of T Pyx was seen
to increase after the 2010 eruption (Patterson et al., this volume),
the period increase only constrains the combination of ejected mass
and angular momentum loss.  If more than the minimum allowed mass
was ejected, so was angular momentum. So the proposed scenario is
that of a common-envelope interaction during the nova eruption,
resulting in an impulsive angular momentum loss, which drives higher
accretion rate for the ensuing decades.  I suggest that such a
scenario deserves serious, quantitative analysis.

\section{Summary and Conclusions}

Recent RN eruptions have be subjected to intense multi-wavelength
observing campaigns using advanced facilities, including the Karl
G. Jansky VLA, many ground-based optical/IR telescopes, HST,
{\sl Swift\/} and other X-ray observatories. Although not a new
discovery as such, recent images of nova ejecta demonstrates once
again that they are not spherically symmetric.  In the case of T Pyx,
the multi-wavelength data strongly suggest that there was a initial,
small ejection and a much larger, delayed ejection.  These facts
together suggest that the binary companion, via common-envelope
interaction, may be involved in the ejection process.

It is expected that RNe harbor massive white dwarfs accreting at
a very high rate. While the accretion rate may be high in some subset
of RNe for evolutionary reasons (symbiotic RNe, and long-period systems),
this is definitely not the case for T~Pyx and IM~Nor.  Unfortunately,
there is a serious difficulty with the commonly invoked mechanism of
irradiation, when the irradiation is by soft X-ray and EUV photons.
I presented a possible scenario involving impulsive angular momentum
loss during the common envelope phase.

\thanks
I thank my colleagues in the E-nova collaboration, particularly
Drs. Nelson, Sokoloski, and Chomiuk, for stimulating discussion.

\bigskip
\bigskip
\noindent {\bf DISCUSSION}

\bigskip
\noindent {\bf CHRISTIAN KNIGGE:} Regarding the possibility that irradiation
might enhance the mass accretion rate: I think that an enhancement
{\sl is\/} possible, at least via irradiation-driven winds from the donor.
In fact, this model was first proposed for SSS, where the irradiation is
all in the soft X-ray/EUV regime.

\bigskip
\noindent {\bf KOJI MUKAI:} It is true that the viability of irradiation-driven
wind to enhance the accretion rate is a separate issue.  I feel that
sometimes the direct effect of irradiation is often too casually
invoked as the explanation, and my intent was to advise caution.

\end{multicols}
\end{document}